# Cargo transport by several motors

Yunxin Zhang*

*Shanghai Key Laboratory for Contemporary Applied Mathematics,*

*Centre for Computational System Biology,*

*School of Mathematical Sciences, Fudan University, Shanghai 200433, China.*

In cells, organelles and vesicles are usually transported by cooperation of several motor proteins, including plus-end directed motor kinesin and minus-end directed motor dynein. Many biophysical models have been constructed to understand the mechanism of this motion. However, so far, the basic principle about it remains unclosed. In this paper, based on the recent experimental results and existing theoretical models, a spider-like model is provided. In this model, each motor is regarded as a bead-spring system. The bead can bind to or unbind from the track stochastically, and step forward or backward with fixed step size $L$ and force dependent transition rates. The spring connects the bead to cargo tightly. At any time, the position of cargo is determined by force balance condition. The obvious characteristics of our model are that, the motors interact with each other and they do *not* share the external load equally. Our results indicate, the stall force of cargo, under which the mean velocity of cargo vanishes, usually decreases with the interactions between motors. If the cargo is pulled by several motors from same motor species, the stall force of cargo is bigger than that of the single motor, but usually smaller than their sum. However, if the cargo is pulled by two motors with different directionality, the stall force of cargo might bigger or smaller than the difference of the two single motor stall forces. The results imply, cooperation of several motors is helpful to pull big cargoes, though it is not so good as might be expected and no obvious help to improve the velocity.

## I. INTRODUCTION

Intracellular transport along microtubule (MT) is usually powered by motor proteins, which transform the free energy released from ATP hydrolysis into mechanical work [1–4], such as conventional kinesin [5–8], which moves to the plus end of MT, and cytoplasmic dynein [9–12], which moves to the minus end of MT. Although the single motor protein properties have been studied in detail both experimentally and theoretically, recent experimental data find the cargoes in cells are usually transport by several motor proteins [13–17], and the stall force and mean velocity of the cargo might completely different from which of the single motor protein [16, 18–20]. Meanwhile, the cooperation of several motor proteins in the motion of muscle [2, 21–25] and flagellar filaments [26–32] have also been observed.

To understand the mechanism of the motion of cargo along MT, many theoretical models have been constructed [15, 33–36]. In [36, 37], Lipowsky and coworkers provided a tug-of-war model, in which the cargo is assumed to be attached by several plus-end directed and minus-end directed motors, and the cargo motion are determined by the number of motors which engage in the tug-of-war. By this model many valuable properties of the intracellular transport of the cargoes can be found [19, 20, 37, 38]. However, there is no evidence that all the binding sites of plus-end directed motors (or minus-end directed motors) on MT have equal distance to the cargo, though experiments found that there are usually several (12-15) parallel tracks along MT [2, 39–42]. So, the load felt by each motors, and consequently the rates of binding to and unbinding from the track of each motors, are different. To account this difference, Kunwar and collaborators regard each motor as a spring in their model, and found that, in fact, the motors can influence each other through the cargo [18]. Recent experimental data indicate that the inter-motor interactions should be taken into account in the modeling of cargo motion [43, 44]. Other valuable discussions about this problem can be found in the literatures [23, 26, 33, 45–49].

*Email: xyz@fudan.edu.cn

In this paper, based on the existing results, a new kinetic model will be presented. In this model, each motor protein is regarded as a bead-spring system. The bead can move forward and backward along the track with step size $L$ ($L = 8$ nm for kinesin and dyenin), and unbind from or bind to the track stochastically. The cargo is connected to each bead by a spring. Roughly speaking, the cargo moves along microtubule like a spider. The position of cargo is obtained by force balance condition, and its velocity depends on the external load and distance between any two of the motor proteins. Using the similar methods as Fisher and Kolomeisky [5, 50, 51], explicit expressions of the mean velocity for some special cases can be obtained. Essentially, in our model, the interactions among motor proteins are described by the similar method as Kunwar *et al* [18], but the motion of single motor protein is modeled by similar idea as Fisher and Kolomeisky [5, 50, 51]. The advantage of our model is that, it not only takes into account the inter-motor interactions explicitly, but also models the single motor in a simple kinetic framework. So, some explicit results can be obtained. Or, in other words, our model is not only more reasonable to describe the motion of cargo pulled by several motors, but also theoretically tractable to get explicit results.

The organization of this paper is as follows. In the next section, the model will be described generally, and then in section **III**, special case, in which there are only two motor proteins, are further studied and explicit results are obtained. In section **IV**, examples of Monte Carlo simulations and theoretical results are given to illustrate the properties of the model. Finally, concluding and remarks are presented in the last section.

## II. MODEL DESCRIPTION

We assume that the cargo is tightly attached by $N$ motor proteins, each of them consists of a bead and a spring. Each bead can jump forward or backward with rates $u_n, w_n$ $(1 \leq n \leq N)$, and step size $L$. The cargo is connected to bead-n by a spring with spring constant $\kappa_n$ [18] [62] At any time, if bead-n is binding to the $l_n-$th binding site of MT, and the external load is $F_{ext}$ (which is positive if it points to the minus-end of MT), then the location $x$ of cargo satisfies the following force balance condition

$$\sum_{j=1}^{N}(l_jL - x)\kappa_j = F_{ext}. \tag{1}$$

Therefore,

$$x = \frac{1}{\sum_{j=1}^{N}\kappa_j}\left(\sum_{j=1}^{N}\kappa_j l_j L - F_{ext}\right). \tag{2}$$

Specially, if $\kappa_n \equiv \kappa$ for $1 \leq n \leq N$, then

$$x = \frac{1}{N}\left(\sum_{j=1}^{N} l_j L - \frac{F_{ext}}{\kappa}\right). \tag{3}$$

If bead-i is unbound from MT, then $l_i L = x$. So, if there are $M$ motors which are unbound from the track, the $N$ in the above formulation should be replaced by $N - M$.

By the expression in Eq. (3), one finds that, if bead-n makes a forward step, i.e., moves from location $l_n L$ to $(l_n + 1)L$, the location of cargo will change from $x$ to $x + \Delta_n L$ with

$\Delta_n = \kappa_n \Big/ \sum_{i=1}^{N} \kappa_i$. So the change of potential of bead-i (for $i \neq n$) is

$$\begin{aligned}\Delta G_i^f(l_n) :=& G_i(l_1, \cdots, l_{n-1}, l_n + 1, l_{n+1}, \cdots, l_N) - G_i(l_1, \cdots, l_N) \\ =& \frac{1}{2}\kappa_i \left[(l_i L - (x + \Delta_n L))^2 - (l_i L - x)^2\right] \\ =& \kappa_i \Delta_n L \left(x - l_i L + \frac{1}{2}\Delta_n L\right) \\ =& \Delta_i \Delta_n L \left(\sum_{j=1}^{N} \kappa_j l_j L - F_{ext}\right) + \kappa_i \Delta_n \left(\frac{1}{2}\Delta_n - l_i\right) L^2,\end{aligned} \tag{4}$$

where $G_i(l_1, \cdots, l_N)$ is the potential of bead-i at state $(l_1, \cdots, l_N)$. And the potential change of bead-n is

$$\begin{aligned}\Delta G_n^f(l_n) :=& G_n(l_1, \cdots, l_{n-1}, l_n + 1, l_{n+1}, \cdots, l_N) - G_n(l_1, \cdots, l_N) \\ =& \frac{1}{2}\kappa_n \left[((l_n + 1)L - (x + \Delta_n L))^2 - (l_n L - x)^2\right] \\ =& \kappa_n (L - \Delta_n L) \left(l_n L - x + \frac{L}{2} - \frac{1}{2}\Delta_n L\right) \\ =& \kappa_n (1 - \Delta_n) \left(l_n + \frac{1}{2} - \frac{1}{2}\Delta_n\right) L^2 - \Delta_n (1 - \Delta_n) \left(\sum_{j=1}^{N} \kappa_j l_j L - F_{ext}\right) L.\end{aligned} \tag{5}$$

So the total potential difference of the system during one forward step of bead-n is

$$\Delta G^f(l_n) = \frac{1}{2}\kappa_n (1 - \Delta_n) L^2 + \Delta_n L^2 \sum_{i=1}^{N} \kappa_i (l_n - l_i) + F_{ext} \Delta_n L. \tag{6}$$

Similarly, if bead-n makes a backward step, i.e., moves from location $l_n L$ to $(l_n - 1)L$, then the location of the cargo will change from $x$ to $x - \Delta_n L$. So the potential change of bead-i (for $i \neq n$) is

$$\begin{aligned}\Delta G_i^b(l_n) :=& G_i(l_1, \cdots, l_N) - G_i(l_1, \cdots, l_{n-1}, l_n - 1, l_{n+1}, \cdots, l_N) \\ =& \Delta_i \Delta_n L \left(\sum_{j=1}^{N} \kappa_j l_j L - F_{ext}\right) - \kappa_i \Delta_n \left(\frac{1}{2}\Delta_n + l_i\right) L^2,\end{aligned} \tag{7}$$

and the change of potential of bead-n is

$$\begin{aligned}\Delta G_n^b(l_n) :=& G_n(l_1, \cdots, l_N) - G_i(l_1, \cdots, l_{n-1}, l_n - 1, l_{n+1}, \cdots, l_N) \\ =& \kappa_n (1 - \Delta_n) \left(l_n - \frac{1}{2} + \frac{1}{2}\Delta_n\right) L^2 - \Delta_n (1 - \Delta_n) \left(\sum_{j=1}^{N} \kappa_j l_j L - F_{ext}\right) L.\end{aligned} \tag{8}$$

Therefore, the total potential difference during one backward step of bead-n is

$$\Delta G^b(l_n) = \frac{1}{2}\kappa_n(\Delta_n - 1)L^2 + \Delta_n L^2 \sum_{i=1}^{N} \kappa_i(l_n - l_i) + F_{ext}\Delta_n L. \tag{9}$$

Using the same as in [5, 51, 52], the forward and backward transition rates of bead-j can be obtained as follows

$$u_j = u_j^0 \exp\left(-\frac{\delta_j \Delta G^f(l_j)}{k_B T}\right), \qquad w_j = \; w_j^0 \exp\left(\frac{(1-\delta_j)\Delta G^b(l_j)}{k_B T}\right), \tag{10}$$

where $u_j^0, w_j^0$ are zero-load forward and backward transition rates respectively, and $\delta_j$ is load distribution factor.

Similar as in [20, 36–38], the rates of binding to and unbinding from the track of bead-j can be obtained as follows

$$k_j^{\text{off}}(l_j) = k_j^{\text{off}} \exp\left(\frac{\kappa_j |l_j L - x|}{F_d^j}\right), \qquad k_j^{\text{on}}(l_j) = k_j^{\text{on}} \exp\left(-\frac{\kappa_j |l_j L - x|}{F_a^j}\right), \tag{11}$$

in which $x$ is the location of cargo (see Eq. (2)), $F_d^j$ is *detachment force*, and we call $F_a^j$ *attachment force* which is assumed to be infinity in the tug-of-war model [36].

## III. CARGO TRANSPORT BY TWO MOTORS

To better understand the basic properties of cargo transport by several motors, in this section, we give detailed discussion about one special case in which there are only two motors, named by motor-1 and motor-2 respectively. Let $l_+L$ be the location of bead-1, $l_-L$ be the location of bead-2, then $lL = (l_+ - l_-)L$ is the distance between the two motors. Let $P(l,t)$ denote the probability of finding the distance between the two motors is $lL$ at time $t$, $P_+(t)$ and $P_-(t)$ be probabilities of finding bead-1 and bead-2 being detached from the track at time $t$ respectively. Then, for this special case, Eq. (1) reduces to

$$\kappa_+(l_+L - x) + \kappa_-(l_-L - x) = F_{ext}. \tag{12}$$

Consequently,

$$x = \frac{\kappa_+ L}{\kappa_+ + \kappa_-} l_+ + \frac{\kappa_- L}{\kappa_+ + \kappa_-} l_- - \frac{F_{ext}}{\kappa_+ + \kappa_-}. \tag{13}$$

It can be verified that the potential at state $(l_+, l_-)$ is

$$\begin{aligned} G(l_+, l_-) =& \frac{1}{(\kappa_+ + \kappa_-)^2} \left[ \kappa_+ \left( \kappa_-(l_+ - l_-)L + F_{ext} \right)^2 + \kappa_- \left( \kappa_+(l_- - l_+)L + F_{ext} \right)^2 \right] \\ &+ \left( \frac{\kappa_+ L}{\kappa_+ + \kappa_-} l_+ + \frac{\kappa_- L}{\kappa_+ + \kappa_-} l_- - \frac{F_{ext}}{\kappa_+ + \kappa_-} \right) F_{ext}. \end{aligned} \tag{14}$$

So

$$\Delta G_+^f := G(l_+ + 1, l_-) - G(l_+, l_-) = \frac{\kappa_+ \kappa_- L^2}{\kappa_+ + \kappa_-} \left( l + \frac{1}{2} \right) + \frac{\kappa_+ L}{\kappa_+ + \kappa_-} F_{ext}. \tag{15}$$

Similarly,

$$\Delta G_-^f := G(l_+, l_- + 1) - G(l_+, l_-) = \frac{\kappa_+ \kappa_- L^2}{\kappa_+ + \kappa_-} \left( \frac{1}{2} - l \right) + \frac{\kappa_- L}{\kappa_+ + \kappa_-} F_{ext}, \tag{16}$$

$$\Delta G_+^b := G(l_+, l_-) - G(l_+ - 1, l_-) = \frac{\kappa_+ \kappa_- L^2}{\kappa_+ + \kappa_-} \left( l - \frac{1}{2} \right) + \frac{\kappa_+ L}{\kappa_+ + \kappa_-} F_{ext}, \tag{17}$$

$$\Delta G_-^b := G(l_+, l_-) - G(l_+, l_- - 1) = -\frac{\kappa_+ \kappa_- L^2}{\kappa_+ + \kappa_-} \left( \frac{1}{2} + l \right) + \frac{\kappa_- L}{\kappa_+ + \kappa_-} F_{ext}. \tag{18}$$

Therefore, $P(l,t), P_+(t), P_-(t)$ are governed by the following master equations [63]

$$\begin{aligned} \frac{d}{dt} P(l,t) =& U(l-1)P(l-1,t) + W(l+1)P(l+1,t) + A_+(l)P_+(t) + A_-(l)P_-(t), \\ & - (U(l) + W(l) + D_+(l) + D_-(l))P(l,t), \quad l = 0, \pm 1, \pm 2, \cdots, \\ \frac{d}{dt} P_+(t) =& \sum_{l=-\infty}^{+\infty} D_+(l)P(l,t) - \sum_{l=-\infty}^{+\infty} A_+(l)P_+(t), \\ \frac{d}{dt} P_-(t) =& \sum_{l=-\infty}^{+\infty} D_-(l)P(l,t) - \sum_{l=-\infty}^{+\infty} A_-(l)P_-(t), \end{aligned} \tag{19}$$

where

$$U(l) = U_+(l) + W_-(l) = U_+ \exp\left( -\frac{\delta_+ \Delta G_+^f}{k_B T} \right) + W_- \exp\left( \frac{(1-\delta_-)\Delta G_-^b}{k_B T} \right), \tag{20}$$

$$W(l) = W_+(l) + U_-(l) = W_+ \exp\left( \frac{(1-\delta_+)\Delta G_+^b}{k_B T} \right) + U_- \exp\left( -\frac{\delta_- \Delta G_-^f}{k_B T} \right), \tag{21}$$

$$D_+(l) = k_{0+}^{\rm off} \exp\left(\frac{\kappa_+|l_+L - x|}{F_d^+}\right) = k_{0+}^{\rm off} \exp\left(\frac{\kappa_+|\kappa_- lL + F_{ext}|}{(\kappa_+ + \kappa_-)F_d^+}\right), \tag{22}$$

$$D_-(l) = k_{0-}^{\rm off} \exp\left(\frac{\kappa_-|l_-L - x|}{F_d^-}\right) = k_{0-}^{\rm off} \exp\left(\frac{\kappa_-|F_{ext} - \kappa_+ lL|}{(\kappa_+ + \kappa_-)F_d^-}\right), \tag{23}$$

$$A_+(l) = k_{0+}^{\rm on} \exp\left(-\frac{\kappa_+|l_+L - x|}{F_a^+}\right) = k_{0+}^{\rm on} \exp\left(-\frac{\kappa_+|\kappa_- lL + F_{ext}|}{(\kappa_+ + \kappa_-)F_a^+}\right), \tag{24}$$

$$A_-(l) = k_{0-}^{\rm on} \exp\left(-\frac{\kappa_-|l_-L - x|}{F_a^-}\right) = k_{0-}^{\rm on} \exp\left(-\frac{\kappa_-|F_{ext} - \kappa_+ lL|}{(\kappa_+ + \kappa_-)F_a^-}\right), \tag{25}$$

where $U_\pm, W_\pm, k_{0\pm}^{\rm off}, k_{0\pm}^{\rm on}, F_d^\pm, F_a^\pm$ are single motor parameters. And $U_+(l), W_+(l)$ are the forward and backward transition rates of bead-1 when $l_+ - l_- = l$, $U_-(l), W_-(l)$ are the forward and backward transition rates of bead-2 when $l_+ - l_- = l$. $A_\pm(l), D_\pm(l)$ are binding and unbinding rates of single motor [15, 36]. Formulations (20-25) can be replaced by more reasonable ones to describe the force dependence of single motor transition rates [5, 53–56], but the basic properties will not change essentially.

From Eq. (13), one sees that, for one step of motor-1 or motor-2, the corresponding displacement of cargo is $\frac{\kappa_+ L}{\kappa_+ + \kappa_-}$ or $\frac{\kappa_- L}{\kappa_+ + \kappa_-}$ respectively. One can also verify that, if bead-1 unbinds from state $l$, the change of cargo position is

$$\frac{\kappa_+ L}{\kappa_+ + \kappa_-}\left(\frac{F_{ext}}{\kappa_- L} + l\right), \tag{26}$$

and if motor-2 unbinds from state $l$, the change of cargo position is

$$\frac{\kappa_- L}{\kappa_+ + \kappa_-}\left(\frac{F_{ext}}{\kappa_+ L} - l\right). \tag{27}$$

So the steady state velocity of the cargo is (see [57])

$$\begin{aligned} V =& \frac{L}{\kappa_+ + \kappa_-} \sum_{l=-\infty}^{\infty} \left[ U_+(l)\kappa_+ + U_-(l)\kappa_- - W_+(l)\kappa_+ - W_-(l)\kappa_- \right. \\ & \left. - \kappa_+ \left( \frac{F_{ext}}{\kappa_- L} + l \right) D_+(l) - \kappa_- \left( \frac{F_{ext}}{\kappa_+ L} - l \right) D_-(l) \right] P(l) \\ & + L \left[ U_- \exp\left( -\frac{\delta_- F_{ext} L}{k_B T} \right) - W_- \exp\left( \frac{(1-\delta_-) F_{ext} L}{k_B T} \right) \right. \\ & \left. + \frac{\kappa_+}{\kappa_+ + \kappa_-} \sum_{l=-\infty}^{\infty} \left( \frac{F_{ext}}{\kappa_- L} + l \right) A_+(l) \right] P_+ \\ & + L \left[ U_+ \exp\left( -\frac{\delta_+ F_{ext} L}{k_B T} \right) - W_+ \exp\left( \frac{(1-\delta_+) F_{ext} L}{k_B T} \right) \right. \\ & \left. + \frac{\kappa_-}{\kappa_+ + \kappa_-} \sum_{l=-\infty}^{\infty} \left( \frac{F_{ext}}{\kappa_+ L} - l \right) A_-(l) \right] P_-, \end{aligned} \tag{28}$$

where the steady state probabilities $P(l), P_+, P_-$ are obtained by Eq. (19).

### A. Special case I

Under the assumption that $A_\pm(l) = D_\pm(l) = 0$ for $l \neq 0$ (see Fig. 1), the steady state master equations are

$$\begin{aligned} & U(l-1)P(l-1) + W(l+1)P(l+1) - (U(l) + W(l))P(l) = 0, \quad l = \pm 1, \pm 2, \cdots, \\ & U(-1)P(-1) + W(1)P(1) + A_+(0)P_+ + A_-(0)P_- - (U(0) + W(0) + D_+(0) + D_-(0))P(0) = 0, \\ & D_+(0)P(0) - A_+(0)P_+ = 0, \\ & D_-(0)P(0) - A_-(0)P_- = 0. \end{aligned} \tag{29}$$

If there is no external load, $A_\pm(l) = 0$ for $l \neq 0$ means the unbinding motor only can rebind to binding site of the track with lowest energy barrier, and the cargo does not change its location during the binding process. $D_\pm(l) = 0$ for $l \neq 0$ means the motor only can unbind from the track when both of them locate at the same position (but not the same binding site, since they might walk along different filaments).

It is reasonable to assume that $P(\pm\infty) = 0$ for $\kappa > 0$, then, for $l > 0$

$$\begin{aligned} P(-l) =& \frac{W(-l+1)}{U(-l)} P(-l+1) = \frac{W(-l+1)\cdots W(0)}{U(-l)\cdots U(-1)} P(0) =: c_{-l} P(0), \\ P(l) =& \frac{U(l-1)}{W(l)} P(l-1) = \frac{U(l-1)\cdots U(0)}{W(l)\cdots W(1)} P(0) =: c_l P(0), \end{aligned} \tag{30}$$

and

$$P_+ = \frac{D_+(0)}{A_+(0)} P(0), \quad P_- = \frac{D_-(0)}{A_-(0)} P(0) \tag{31}$$

So the normalization condition $P_+ + P_- + \sum_{l=-\infty}^{\infty} P(l) = 1$ gives

$$P(0) = \frac{1}{D_+(0)/A_+(0) + D_-(0)/A_-(0) + \sum_{l=-\infty}^{\infty} c_l}, \tag{32}$$

in which $c_0 = 1$. Consequently,

$$\begin{aligned} P(l) =& \frac{c_l}{D_+(0)/A_+(0) + D_-(0)/A_-(0) + \sum_{l=-\infty}^{\infty} c_l} \\ =& \begin{cases} \dfrac{\prod_{i=1}^{l} \frac{U(i-1)}{W(i)}}{1 + D_+(0)/A_+(0) + D_-(0)/A_-(0) + \sum_{l=1}^{\infty} \prod_{i=1}^{l} \frac{U(i-1)}{W(i)} + \sum_{l=-1}^{-\infty} \prod_{i=l}^{-1} \frac{W(i+1)}{U(i)}}, & \text{for } l \geq 1, \\ \dfrac{\prod_{i=l}^{-1} \frac{W(i+1)}{U(i)}}{1 + D_+(0)/A_+(0) + D_-(0)/A_-(0) + \sum_{l=1}^{\infty} \prod_{i=1}^{l} \frac{U(i-1)}{W(i)} + \sum_{l=-1}^{-\infty} \prod_{i=l}^{-1} \frac{W(i+1)}{U(i)}}, & \text{for } l \leq -1. \end{cases} \end{aligned} \tag{33}$$

One can verify that, the steady state mean velocity of the cargo is (see (28))

$$\begin{aligned} V =& \frac{L}{\kappa_+ + \kappa_-} \sum_{l=-\infty}^{\infty} \left[ U_+(l)\kappa_+ + U_-(l)\kappa_- - W_+(l)\kappa_+ - W_-(l)\kappa_- \right] P(l) \\ &+ L \left[ U_- \exp\left( -\frac{\delta_- F_{ext} L}{k_B T} \right) - W_- \exp\left( \frac{(1-\delta_-) F_{ext} L}{k_B T} \right) \right] P_+ \\ &+ L \left[ U_+ \exp\left( -\frac{\delta_+ F_{ext} L}{k_B T} \right) - W_+ \exp\left( \frac{(1-\delta_+) F_{ext} L}{k_B T} \right) \right] P_- \\ &- \left( \frac{1}{\kappa_-} - \frac{1}{\kappa_+ + \kappa_-} \right) F_{ext} D_+(0) P(0) - \left( \frac{1}{\kappa_+} - \frac{1}{\kappa_+ + \kappa_-} \right) F_{ext} D_-(0) P(0) \\ &+ \left( \frac{1}{\kappa_-} - \frac{1}{\kappa_+ + \kappa_-} \right) F_{ext} A_+(0) P_+ - \left( \frac{1}{\kappa_+} - \frac{1}{\kappa_+ + \kappa_-} \right) F_{ext} A_-(0) P_-. \end{aligned} \tag{34}$$

### B. Special case II

Another meaningful special case is that, $P(l) = 0$ for $|l| > n$, and the motor only can unbind from the track at states $l = \pm n$ and rebind to the track at state $l = 0$ (see Fig. 2). This means the unbinding motors only can rebind to binding site of the track with lowest energy barrier, and unbind from the binding site with strongest inter-motor interaction. From the steady state master equations one easily sees

$$\begin{aligned}
&U(l-1)P(l-1) - W(l)P(l) = J_-, \qquad \text{for} \;\; l = -n+1, \cdots, 0,\\
&U(l-1)P(l-1) - W(l)P(l) = J_+, \qquad \text{for} \;\; l = 1, \cdots, n,\\
&D_+(-n)P(-n) + D_-(-n)P(-n) = J_-,\\
&D_+(n)P(n) + D_-(n)P(n) = J_+,\\
&D_+(-n)P(-n) + D_+(n)P(n) = A_+(0)P_+,\\
&D_-(-n)P(-n) + D_-(n)P(n) = A_-(0)P_-,\\
&A_+(0)P_+ + A_-(0)P_- = J_+ - J-,
\end{aligned} \tag{35}$$

where $J_\pm$ are constants. Under the normalization condition the probabilities $P(l)$ can be obtained by the above equations, and then the mean velocity can be obtained by (see (28))

$$\begin{aligned}
V =& \frac{L}{\kappa_+ + \kappa_-} \sum_{l=-\infty}^{\infty} \left[U_+(l)\kappa_+ + U_-(l)\kappa_- - W_+(l)\kappa_+ - W_-(l)\kappa_-\right] P(l)\\
&- \frac{L}{\kappa_+ + \kappa_-} \left[\kappa_+ \left(\frac{F_{ext}}{\kappa_- L} + n\right) D_+(n) - \kappa_- \left(\frac{F_{ext}}{\kappa_+ L} - n\right) D_-(n)\right] P(n)\\
&- \frac{L}{\kappa_+ + \kappa_-} \left[\kappa_+ \left(\frac{F_{ext}}{\kappa_- L} - n\right) D_+(-n) - \kappa_- \left(\frac{F_{ext}}{\kappa_+ L} + n\right) D_-(-n)\right] P(-n)\\
&+ \left(\frac{1}{\kappa_-} - \frac{1}{\kappa_- + \kappa_+}\right) F_{ext} A_+(0) P_+ + \left(\frac{1}{\kappa_+} - \frac{1}{\kappa_- + \kappa_+}\right) F_{ext} A_-(0) P_-\\
&+ L \left[U_- \exp\left(-\frac{\delta_- F_{ext} L}{k_B T}\right) - W_- \exp\left(\frac{(1-\delta_-) F_{ext} L}{k_B T}\right)\right] P_+\\
&+ L \left[U_+ \exp\left(-\frac{\delta_+ F_{ext} L}{k_B T}\right) - W_+ \exp\left(\frac{(1-\delta_+) F_{ext} L}{k_B T}\right)\right] P_-.
\end{aligned} \tag{36}$$

## IV. EXAMPLES

In this section, we will give some examples to illustrate the properties of cargo transport by two motors. In all examples, $L = 8$ nm, $\delta_\pm = 0.5$, and $k_BT = 4.12$ pN·nm are used.

Firstly, we use Monte Carlo simulations to observe the distance $l = l_+ - l_-$ between the two motors. In the simulations, we assumed that the motors cannot unbind from the track. From Figs. 3 and 4, we can find that, if the two motors are the same, the value of $l$ fluctuates around 0, otherwise, the average value of $l = l_+ - l_-$ might not be zero. However, in any cases, the amplitude of fluctuation decreases with spring constant $\kappa$. Roughly speaking, if the intrinsic directionality of the two motors are different, the cargo will be moved in an inch worm manner [64] (see Fig. 4). Obviously, the average velocity of the asymmetric cases, as given in Fig. 4, is much lower than that of the symmetric cases, as given in Fig. 3, since there will be a tug-of-war between the two motors if they try to move to opposite directions [14, 36, 58].

In the following, we will study the steady state properties using model (19). From the results in Fig. 5, one finds that, the probability that cargo is transported by only one motor decreases with $F_a^\pm = F_d^\pm$ and $k_{0\pm}^{\text{on}}$, but increases with $k_{0\pm}^{\text{off}}$ (Note, $P_+ + P_- = 1 - \sum_{l=-\infty}^{\infty} P(l)$). Because a large $F_a^\pm = F_d^\pm$ or $k_{0\pm}^{\text{on}}$ means the motor is difficult to detach from the track or more likely to rebind to the track, while a large $k_{0\pm}^{\text{off}}$ means the motor is easy to detach from the track. As have been found in the Monte Carlo simulations, the fluctuation of $l = l_+ - l_-$ decreases with $\kappa$. If the cargo is pulled by two different motors, then, generally $P(l) \neq P(-l)$, which implies that the leading motor and the rear motor don't change their roles frequently. So, in most of the time, the cargo moves in an inch worm manner. From the results in Fig. 6, one sees that, the mean velocity $V$ of cargo decreases with $\kappa$ and external load $F_{ext}$. The stall force $F_s$ of the cargo, i.e., the external load under which $V$ vanishes, is generally different from the sum or difference of that of the two motors, $F_s \neq |F_s^+ \pm F_s^-| := |k_B T[|\ln(U_+/W_+)| \pm |\ln(U_-/W_-)|]/L|$ [5]. If the intrinsic directionality of the two motors are the same, then $F_s$ is smaller than $F_s^+ + F_s^-$, but usually bigger than $F_s^+, F_s^-$ (see Fig. 6(a)(b)). Therefore, cooperation of several motors is helpful for transport of big cargoes. However, if the intrinsic directionality of the two motors are different, it is possible for both $F_s > |F_s^+ - F_s^-|$ and $F_s < |F_s^+ - F_s^-|$ (see Fig. 6(c)(d)). Similar results have been obtained using tug-of-war model [20].

To know the influence of the spring constant $\kappa$ on the motion of cargo, we plotted the stall force $F_s$ and zero-load velocity $V_0$ of the cargo as functions of $\kappa$ in Fig. 7. From the curves (a) (b) in figures about $F_s$, one sees that, if the cargo is pulled by two same motors then, in small $\kappa$ limit, the stall force $F_s$ is two times of the single motor stall forces, but in large $\kappa$ limit, $F_s$ is around the single motor stall force. At the same time, from the curves (a) (b) in figures about $V_0$, one also can see that, for the two same motors cases, the zero-load cargo velocity $V_0$ in small $\kappa$ limit is around the single motor velocity. Meanwhile, the results in Fig. 7 also indicate that, $F_s$ and $V_0$ do *not* always decrease with $\kappa$. In other words, although in most cases, as had been found experimentally by Diehl and coworkers [43], the interaction between two motors is negative to the motion of cargo, in some cases, the increase of this interaction is helpful to improve the performance of the motor team. From the figures, one can also find that, for large $\kappa$, $V_0$ tends to constant. The reason is that, for large $\kappa$, almost all the motion of cargo is made by a single motor (with the other one being detached from the track). Due to the stronge interaction between the two motors, none of them can make a forward or backward step if they are both binding to the track. The difference among the limits of $V_0$ comes from the different values of $P_+$ and $P_-$, which depend on single motor parameters $U_\pm, W_\pm, F_a^\pm, F_d^\pm, k_{0\pm}^{\text{off}}, k_{0\pm}^{\text{on}}$ and $\delta_\pm$.

## V. CONCLUDING REMARKS

In this paper, a theoretical model for the cargo transport by several motors is given. In this model, the location of cargo is determined by force balance condition. Each motor is regarded as a bead-spring system, the bead can bind to or unbind from the track and jump forward or backward with step size $L$, the spring connects the bead to the cargo tightly. With small spring constant, the cargo usually moves with high velocity and has big stall force. Actually, from calculations, we find that, in the small spring constant limit, the velocity of cargo is the average of that of single motors. On the contrary, with big spring constant, the cargo will move with low velocity and has small stall force. If the cargo is pulled by cooperation of two motors with same intrinsic directionality. the stall force of cargo is usually smaller than the sum of stall forces of the two motors, and bigger than each of them. If the motor is transported by two motors with different intrinsic directionality, it is possible that, the stall force of cargo is bigger or smaller than the absolute value of the difference of single motor stall forces. For these cases, the cargo will move in an inch worm manner. Although in most cases, the interaction between the two motors is negative to the motion of cargo, in some cases, the increase of this interaction is helpful to improve the performance of the team.

In our model, the idea to describe interactions among different motors is similar as the one used by Kunwar and coworkers [18], the method to model the motion of single motor is similar as that provided by Fisher and Kolomeisky [5]. So, from this point of view, this method is not only reasonable to describe the motion of organelles and vesicles in cells, which are transported by plus-end directed motor kinesin and minus-end directed motor dynein, but also theoretically tractable to get valuable results. It is also obvious that, the dynamic framework of our model is the same as the one devised by Lipowsky and coworkers [15, 19, 36].

**Acknowledgments**

This work is funded by the National Natural Science Foundation of China (Grant No. 10701029).

---

[1] Mark J. Schnitzer and Steven M. Block. Kinesin hydrolyses one ATP per 8-nm step. *Nature*, 388:386–390, 1997.

[2] J. Howard. *Mechanics of Motor Proteins and the Cytoskeleton.* Sinauer Associates and Sunderland, MA, 2001.

[3] R. D. Vale. The molecular motor toolbox for intracellular transport. *Cell*, 112:467–480, 2003.

[4] A. B. Kolomeisky and M. E. Fisher. Molecular motors: A theorists perspective. *Ann. Rev. Phys. Chem.*, 58:675–695, 2007.

[5] M. E. Fisher and A. B. Kolomeisky. Simple mechanochemistry describes the dynamics of kinesin molecules. *Proc. Natl. Acad. Sci. USA*, 98:7748–7753, 2001.

[6] N. J. Carter and R. A. Cross. Mechanics of the kinesin step. *Nature*, 435:308–312, 2005.

[7] Steven M. Block. Kinesin motor mechanics: Binding, stepping, tracking, gating, and limping. *Biophys. J.*, 92:2986–2995, 2007.

[8] Y. Zhang. Three phase model of the processive motor protein kinesin. *Biophys. Chem.*, 136:19–22, 2008.

[9] S. L. Reck-Peterson, A. Yildiz, A. P. Carter, A. Gennerich, N. Zhang, and R. D. Vale. Single-molecule analysis of dynein processivity and stepping behavior. *Cell*, 126:335–348, 2006.

[10] S. Toba, T. M. Watanabe, L. Yamaguchi-Okimoto, Y. Y. Toyoshima, and H. Higuchi. Overlapping hand-over-hand mechanism of single molecular motility of cytoplasmic dynein. *Proc. Natl. Acad. Sci. USA*, 103:5741–5745, 2006.

[11] Arne Gennerich, Andrew P. Carter, Samara L. Reck-Peterson, and Ronald D. Vale. Force-induced bidirectional stepping of cytoplasmic dynein. *Cell*, 131:952–965, 2007.

[12] A. Houdusse and A. P. Carter. Dynein swings into action. *Cell*, 136:395–396, 2009.

[13] S. Gross, M. Vershinin, and G. Shubeita. Cargo transport: two motors are sometimes better than one. *Curr. Biol.*, 17:R478–R486, 2007.

[14] V. Soppina, A. K. Rai, A. J. Ramaiya, P. Barak, and R. Mallik. Tug-of-war between dissimilar teams of microtubule motors regulates transport and fission of endosomes. *Proc. Natl. Acad. Sci. USA*, 106:19381–19386, 2009.

[15] Stefan Klumpp and Reinhard Lipowsky. Cooperative cargo transport by several molecular motors. *Proc. Natl. Acad. Sci. USA*, 102:17284–17289, 2005.

[16] George T. Shubeita, Susan L. Tran, Jing Xu, Michael Vershinin, Silvia Cermelli, Sean L. Cotton, Michael A. Welte, and Steven P. Gross. Consequences of motor copy number on the intracellular transport of kinesin-1-driven lipid droplets. *Cell*, 135:1098–1107, 2008.

[17] Roop Mallik and Steven P. Gross. Intracellular transport: How do motors work together? *Curr. Biol.*, 19:R416–R418, 2009.

[18] A. Kunwar, M. Vershinin, J. Xu, and S. P. Gross. Stepping, strain gating, and an unexpected force-velocity curve for multiple-motor-based transport. *Curr. Biol.*, 18:1173–1183, 2008.

[19] R. Lipowsky, J. Beeg, R. Dimova, S. Klumpp, and M.J.I. Müller. Cooperative behavior of

molecular motors: Cargo transport and traffic phenomena. *Physica E*, 42:649, 2010.

[20] Y. Zhang and M. E. Fisher. Dynamics of the tug-of-war model for cellular transport. *Phys. Rev. E (in press)*, 2010.

[21] F. Jülicher and J. Prost. Cooperative molecular motors. *Phys. Rev. Lett.*, 75:2618–2621, 1995.

[22] F. Jülicher, A. Ajdari, and J. Prost. Modeling molecular motors. *Rev. Mod. Phys.*, 69:1269, 1997.

[23] M. Badoual, F. Jülicher, and J. Prost. Bidirectional cooperative motion of molecular motors. *Proc. Natl. Acad. Sci. USA*, 99:6696, 2002.

[24] C. Veigel, J. E. Molloy, S.n Schmitz, and J. Kendrick-Jones. Load-dependent kinetics of force production by smooth muscle myosin measured with optical tweezers. *Nat. Cell. Biol.*, 5:980–986, 2003.

[25] Sam Walcott and Sean X. Sun. Hysteresis in cross-bridge models of musclew. *Phys. Chem. Chem. Phys.*, 11:4871–4881, 2009.

[26] T. Elston and G. Oster. Protein turbines I: The bacterial flagellar motor. *Biophys. J.*, 73:703, 1997.

[27] S. Camalet, F. Jülicher, and J. Prost. Self-organized beating and swimming of internally driven filaments. *Phys. Rev. Lett.*, 82:1590, 1999.

[28] Jianhua Xing, Fan Bai, Richard Berry, and George Oster. Torque-speed relationship of the bacterial flagellar motor. *Proc. Natl. Acad. Sci. USA*, 103:1260–1265, 2006.

[29] J. Howard. Mechanical signaling in networks of motor and cytoskeletal proteins. *Annu. Rev. Biophys*, 38:217, 2009.

[30] Giovanni Meacci and Yuhai Tu. Dynamics of the bacterial flagellar motor with multiple stators. *Proc. Natl. Acad. Sci. USA*, 106:3746–3751, 2009.

[31] T. Mora, H. Yu, and N. S. Wingreen. Modeling torque versus speed, shot noise, and rotational diffusion of the bacterial flagellar motor. *Phys. Rev. Lett.*, 103:248102, 2009.

[32] T. J. Mitchison and H. M. Mitchison. How cilia beat. *Nature*, 463:308, 2010.

[33] C. Kural, H. Kim, S. Syed, G. Goshima, V. I. Gelfand, and P. R. Selvin. Kinesin and dynein move a peroxisome in vivo: A tug-of-war or coordinated movement? *Science*, 308:1469, 2005.

[34] X. Nan, P. Chen P. A. Sims an, and X. S. Xie. Observation of individual microtubule motor steps in living cells with endocytosed quantum dots. *J. Phys. Chem. B*, 109:24220, 2005.

[35] V. Levi, A. S. Serpinskaya, E. Gratton, and V. Gelfand. Organelle transport along microtubules in xenopus melanophores: Evidence for cooperation between multiple motors. *Biophys. J.*, 90:318, 2006.

[36] M. J. I. Müller, S. Klumpp, and R. Lipowsky. Tug-of-war as a cooperative mechanism for bidirectional cargo transport by molecular motors. *Proc. Natl. Acad. Sci. USA*, 105:4609–4614, 2008.

[37] Melanie J.I. Müller, Stefan Klumpp, and Reinhard Lipowsky. Motility states of molecular motors engaged in a stochastic tug-of-war. *J. Stat. Phys.*, 133:1059–1081, 2008.

[38] Y. Zhang. Properties of tug-of-war model for cargo transport by molecular motors. *Phys. Rev. E*, 79:061918, 2009.

[39] P. R. Burton and R. H. Himes. Electron microscope studies of pH effects on assembly of tubulin free of associated proteins. *J. Cell. Biol.*, 77:120–133, 1978.

[40] T. Mitchison L. Evans and M. Kirschner. Influence of the centrosome on the structure of nucleated microtubules. *J. Cell. Biol.*, 100:1185–1191, 1985.

[41] D. Chrétien, E. Metoz, E. Verde, E. Karsenti, and R. H. Wade. Lattice defects in microtubules: Protofilament numbers vary within individual microtubules. *J. Cell. Biol.*, 117:1032–1040,

1992.

[42] U. Raviv, T. Nguyen, R. Ghafouri, D. J. Needleman, Y. Li, H. P. Miller, L. Wilson, R. F. Bruinsma, and C. R. Safinya. Microtubule protofilament number is modulated in a stepwise fashion by the charge density of an enveloping layer. *Biophys. J.*, 92:278–287, 2007.

[43] P. E. Constantinou D. K. Jamison A. R. Rogers, J. W. Driver and M. R. Diehl. Negative interference dominates collective transport of kinesin motors in the absence of load. *Phys. Chem.-Chem. Phys.*, 11:4882, 2009.

[44] J.W. Driver, A.R. Rogers, D.K. Jamison, R.K. Das, A.B. Kolomeisky, and M.R. Diehl. Coupling between motor proteins determines dynamic behaviors of motor protein assemblies. *Phys. Chem.-Chem. Phys. (in press)*, 2010.

[45] D. Bray. *Cell movements: from molecules to motility, 2nd Edn.* Garland, New York, 2001.

[46] D. B. Hill, M. J. Plaza, K. Bonin, and G. Holzwarth. Fast vesicle transport in pc12 neurites: velocities and forces. *Eur. Biophys. J.*, 33:623, 2004.

[47] E. B. Stukalin, III H. Phillips, and A. B. Kolomeisky. Coupling of two motor proteins: A new motor can move faster. *Phys. Rev. Lett.*, 94:238101, 2005.

[48] S. W. Grill, K. Kruse, and F. Jülicher. Theory of mitotic spindle oscillations. *Phys. Rev. Lett.*, 94:108104, 2005.

[49] I. H. Riedel-Kruse, A. Hilfinger, J. Howard, and F. Jülicher. How molecular motors shape the flagellar beat. *HFSP J.*, 1:192, 2007.

[50] B. Derrida. Velocity and diffusion constant of a periodic one-dimensional hopping model. *J. Stat. Phys.*, 31:433–450, 1983.

[51] M. E. Fisher and A. B. Kolomeisky. Molecular motors and the forces they exert. *Physica A*, 274:241–266, 1999.

[52] Y. Zhang. A general two-cycle network model of molecular motors. *Physica A*, 383:3465–3474, 2009.

[53] M. E. Fisher and A. B. Kolomeisky. The force exerted by a molecular motor. *Proc. Natl. Acad. Sci. USA*, 96:6597, 1999.

[54] A. B. Kolomeisky and M. E. Fisher. A simple kinetic model describes the processivity of myosin-v. *Biophys. J.*, 84:1642–1650, 2003.

[55] M. E. Fisher and Y. C. Kim. Kinesin crouches to sprint but resists pushing. *Proc. Natl. Acad. Sci. USA*, 102:16209–16214, 2005.

[56] D. Tsygankov and M. E. Fisher. Mechanoenzymes under superstall and large assisting loads reveal structural features. *Proc. Natl. Acad. Sci. USA*, 104:19321–19326, 2007.

[57] Y. Zhang. Derivation of diffusion coefficient of a brownian particle in tilted periodic potential from the coordinate moments. *Physics Letters A*, 373:2629–2633, 2009.

[58] A. Gennerich and D. Schild. Finite-particle tracking reveals submicroscopic-size changes of mitochondria during transport in mitral cell dendrites. *Phys. Biol.*, 3:45–53, 2006.

[59] Ahmet Yildiz, Michio Tomishige, Arne Gennerich, and Ronald D. Vale. Intramolecular strain coordinates kinesin stepping behavior along microtubules. *Cell*, 134:1030–1041, 2008.

[60] Changbong Hyeon and José N. Onuchic. Internal strain regulates the nucleotide binding site of the kinesin leading head. *Proc. Natl. Acad. Sci. USA*, 104:2175–2180, 2007.

[61] C. Bouchiat, M. D. Wang, J.-F. Allemand, T. Strick, S. M. Block, and V. Croquette. Estimating the persistence length of a worm-like chain molecule from force-extension measurements. *Biophys. J.*, 76:409–413, 1999.

[62] For simplicity, in this paper, we assumed that the interactions between beads and cargo can be described by springs. More sophisticated formulations (see [59–61]) can be used here to

model this action, which are used to get cargo position $x$ by motor positions $l_n$ $(1 \leq n \leq N)$. Consequently, more general formulations can be employed in the following discussion to calculate the internal energy in the necker linkers, which is used to determine the transition rates $u_n, w_n$ $(1 \leq n \leq N)$. Moreover, the free length of the springs are assumed to be zero in our analysis.

[63] since we will not study the mean dwell time of cargo on the track and the mean run length of cargo, we assumed in this paper that, the cargo cannot detach from the track. Otherwise, one more process should be added in the model [44], and consequently, two more equations should be added in (19).

[64] In fact, the cargo will move in an inch worm manner if there is a big difference between the velocity of the two motors.

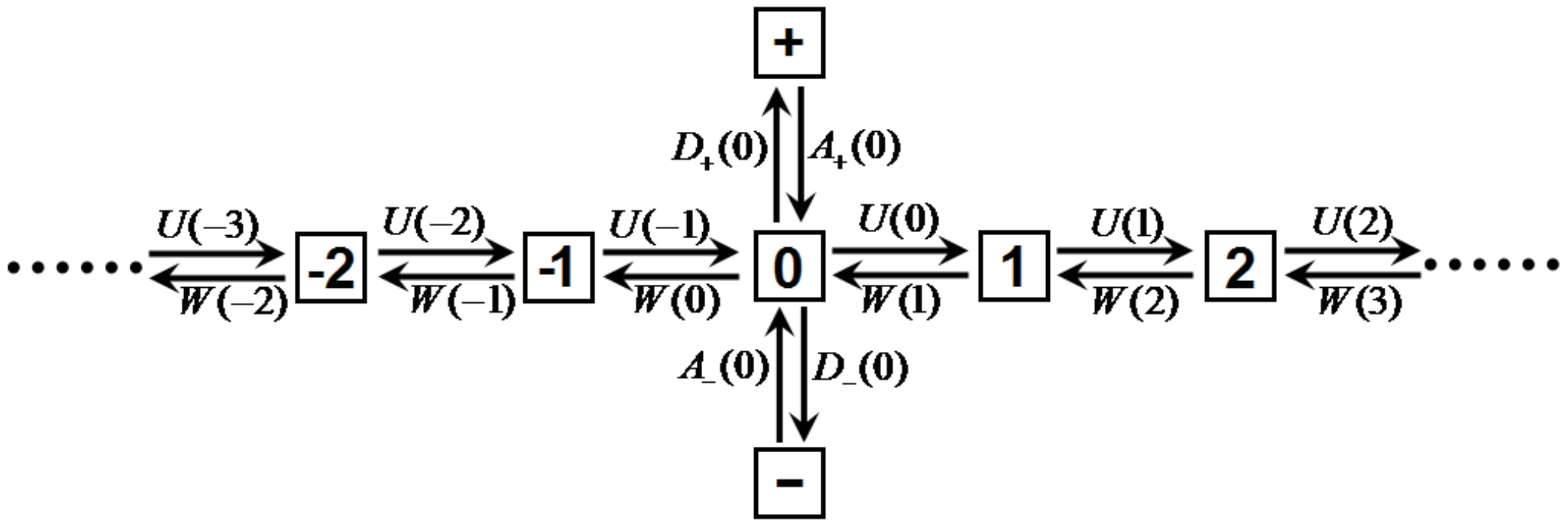

FIG. 1: Schematic depiction of the special case **I**: $A_{\pm}(l) = D_{\pm}(l) = 0$ for $l \neq 0$.

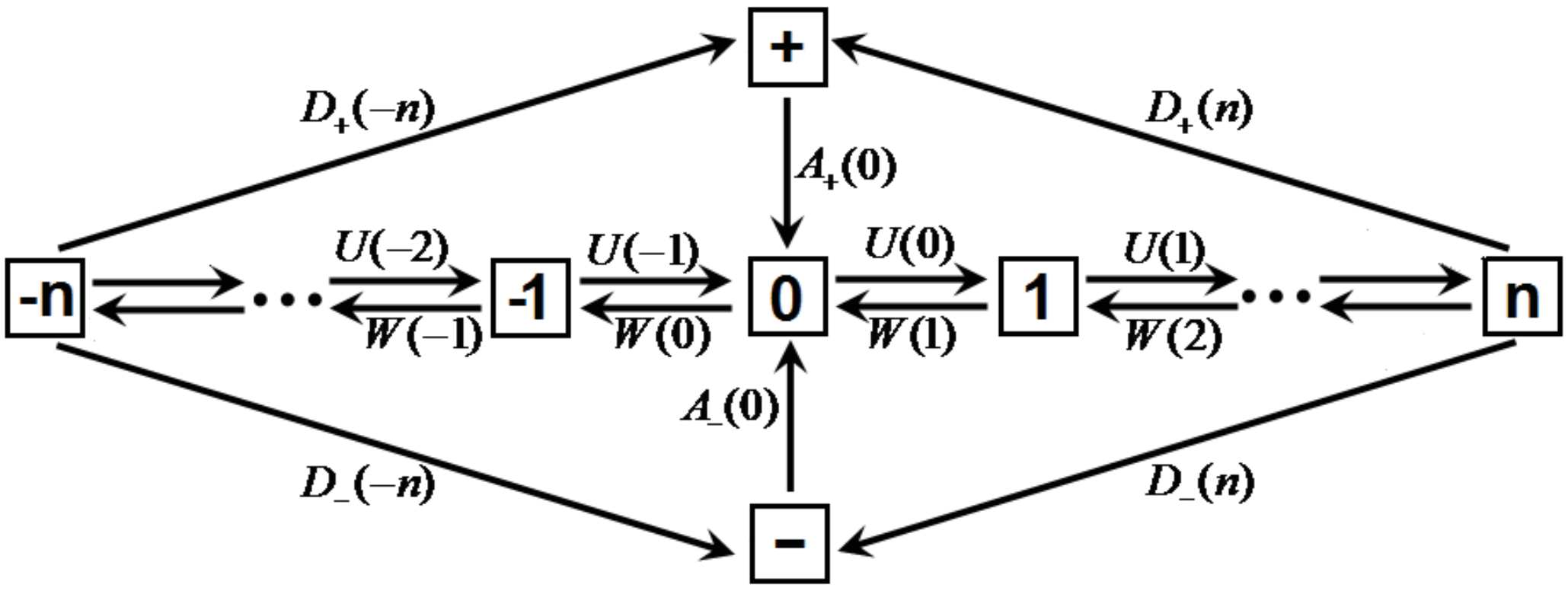

FIG. 2: Schematic depiction of the special case **II**: $P(l) = 0$ for $|l| > n$, and the two motors only can unbind from the track at states $l = \pm n$ and bind to the track at state $l = 0$.

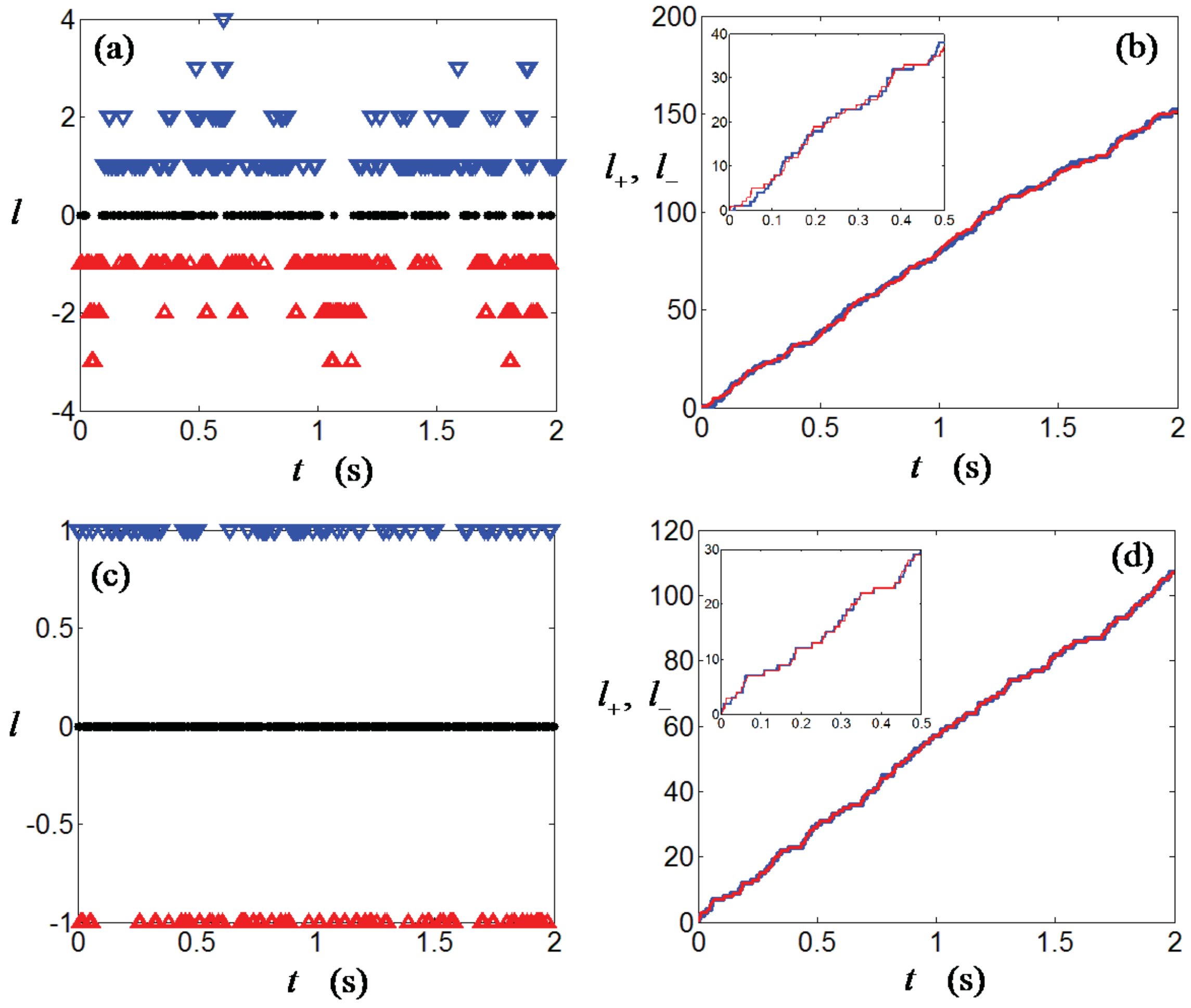

FIG. 3: Monte Carlo simulations of the model with two same motors. In the simulations, it is assumed that the motors cannot unbind from track. The parameters used here are: $U_+ = U_- = 100$ s$^{-1}$, $W_+ = W_- = 1$ s$^{-1}$, $\delta_\pm = 0.5$, $k_BT = 4.12$ pN·nm, $L = 8$ nm, $F_{ext} = 0$ pN, and $\kappa = 0.1$ pN/nm in figures (a) (b), $\kappa = 0.5$ pN/nm in figures (c) (d). The coordinator of the two motors are denoted by $l_+$, $l_-$ respectively, and $l = l_+ - l_-$.

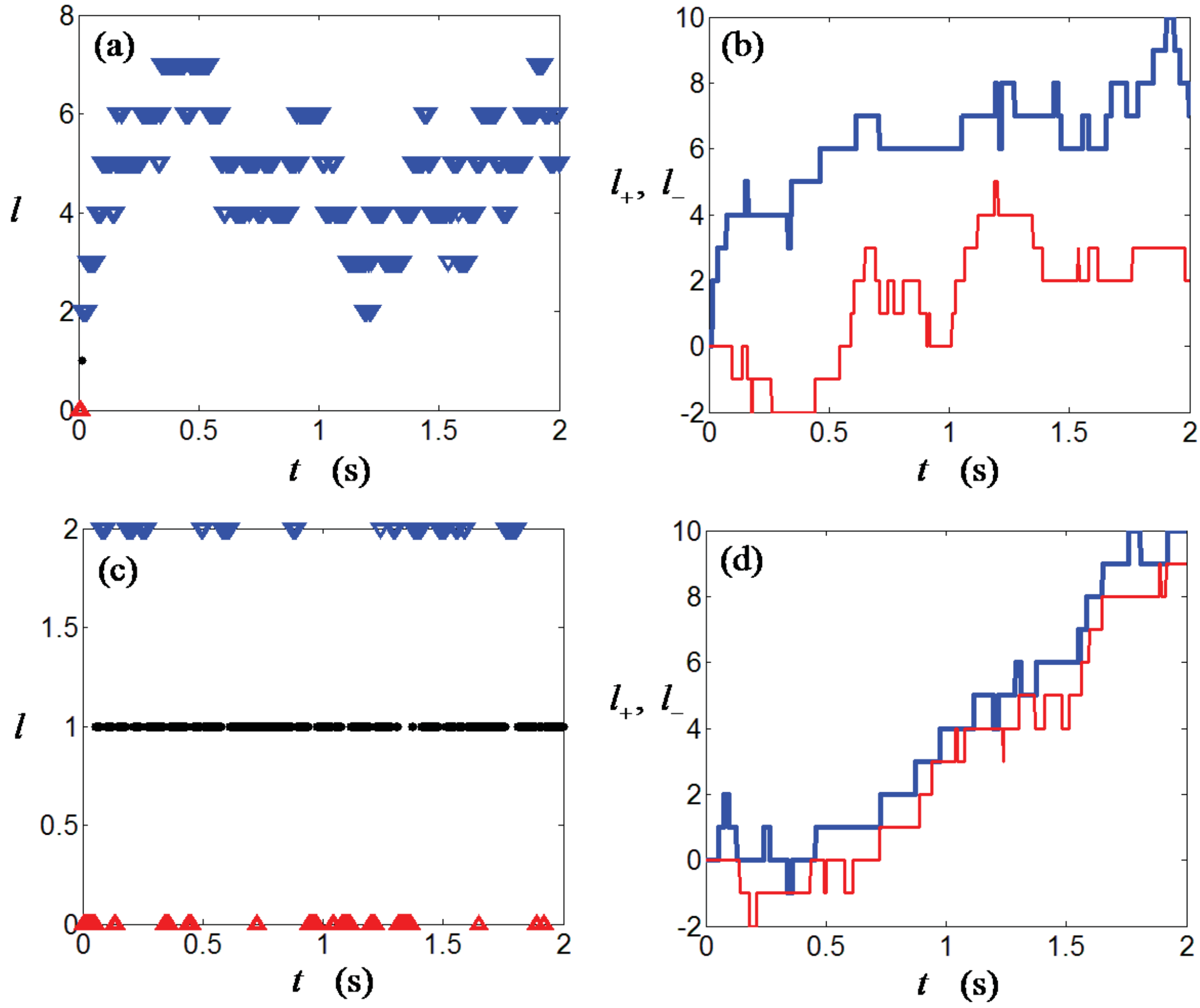

FIG. 4: Monte Carlo simulations of the model with two different motors. The parameters are: $U_+ = 100\ \mathrm{s}^{-1}$, $U_- = 1\ \mathrm{s}^{-1}$, $W_+ = 1\ \mathrm{s}^{-1}$, $W_- = 50\ \mathrm{s}^{-1}$, $\delta_\pm = 0.5$, $k_B T = 4.12$ pN·nm, $L = 8$ nm, $F_{ext} = 0$ pN, and $\kappa = 0.1$ pN/nm in figures (a) (b), $\kappa = 0.5$ pN/nm in figures (c) (d).

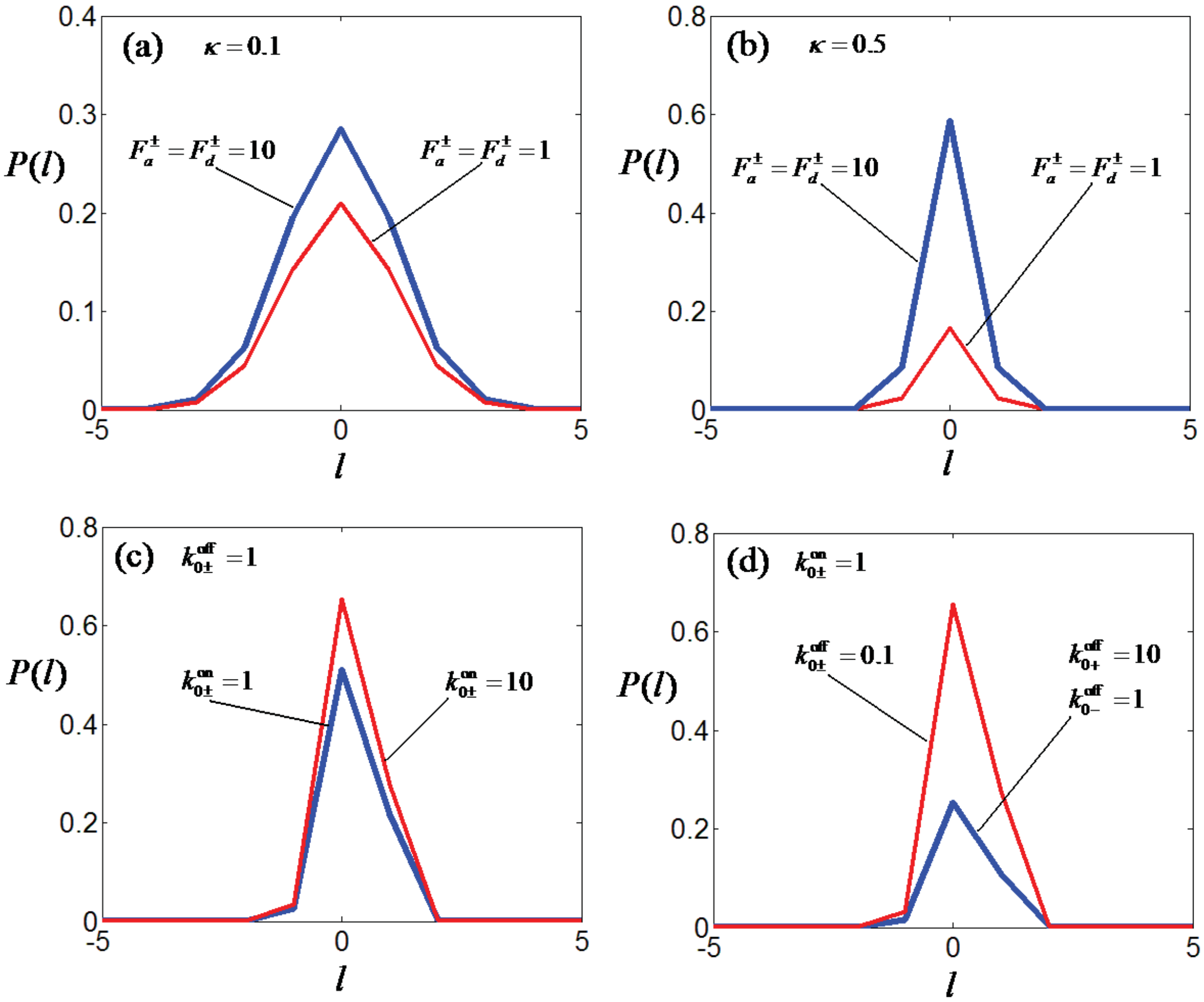

FIG. 5: The steady state probability $P(l)$ of model (19). The parameters used in figures (a) (b) are: $U_+ = U_- = 100$ s$^{-1}$, $W_+ = W_- = 1$ s$^{-1}$, $k^{\rm on}_{0\pm} = k^{\rm off}_{0\pm} = 1$ s$^{-1}$, and $U_+ = 100$ s$^{-1}$, $U_- = 1$ s$^{-1}$, $W_+ = W_- = 50$ s$^{-1}$, $F_a^{\pm} = F_d^{\pm} = 10$ pN, $\kappa = 0.5$ pN/nm in figures (c) (d).

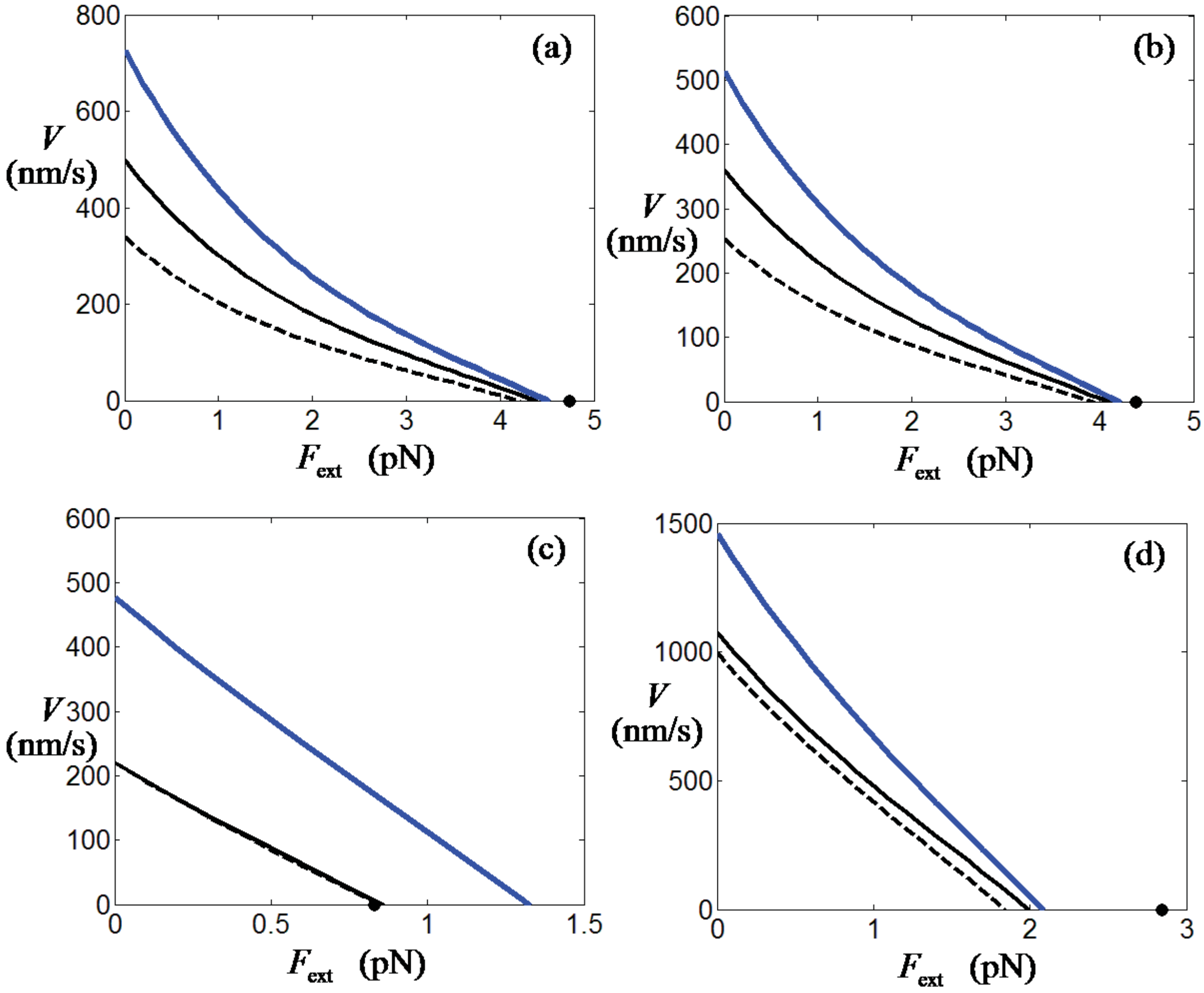

FIG. 6: The velocity force relation for a cargo transport by two motors. The parameters used in the calculations are $k_{0\pm}^{\rm on} = 10$ s$^{-1}$, $k_{0\pm}^{\rm off} = 1$ s$^{-1}$, $F_a^{\pm} = F_d^{\pm} = 10$ pN, and (a) $U_{\pm} = 100$ s$^{-1}$, $W_{\pm} = 1$ s$^{-1}$, (b) $U_+ = 100$ s$^{-1}$, $W_+ = 1$ s$^{-1}$, $U_- = 50$ s$^{-1}$, $W_- = 1$ s$^{-1}$, (c) $U_+ = 1000$ s$^{-1}$, $W_+ = 1$ s$^{-1}$, $U_- = 1$ s$^{-1}$, $W_- = 200$ s$^{-1}$, and (d) $U_+ = 1000$ s$^{-1}$, $W_+ = 1$ s$^{-1}$, $U_- = 50$ s$^{-1}$, $W_- = 200$ s$^{-1}$. The thick lines correspond to $\kappa = 0.1$ pN/nm, the thin lines correspond to $\kappa = 0.5$ pN/nm, and the dashed lines correspond to $\kappa = 0.8$ pN/nm. The black dots are obtained by $k_BT[\ln(U_+/W_+) + \ln(U_-/W_-)]/L$.

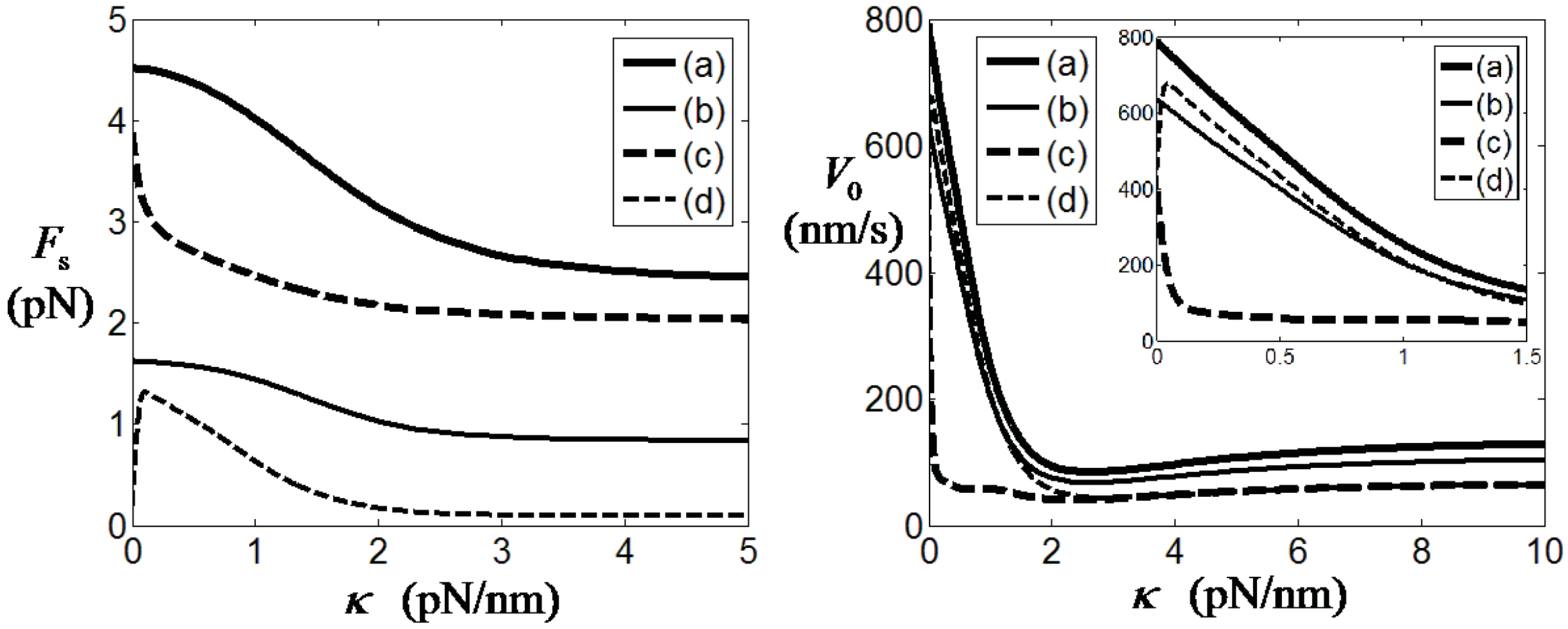

FIG. 7: The stall force $F_s$ (**Left**) and zero-load velocity $V_0$ (**Right**) of cargo transport by two motors as functions of spring constant $\kappa$. The parameters used in the calculations are $k_{0\pm}^{\rm on} = 10$ s$^{-1}$, $k_{0\pm}^{\rm off} = 1$ s$^{-1}$, $F_a^\pm = F_d^\pm = 10$ pN, and (a) $U_\pm = 100$ s$^{-1}$, $W_\pm = 1$ s$^{-1}$, (b) $U_\pm = 100$ s$^{-1}$, $W_\pm = 20$ s$^{-1}$, (c) $U_+ = 100$ s$^{-1}$, $W_+ = 1$ s$^{-1}$, $U_- = W_- = 1$ s$^{-1}$, and (d) $U_+ = 100$ s$^{-1}$, $W_+ = 1$ s$^{-1}$, $U_- = W_- = 500$ s$^{-1}$. The single motor stall forces are: (a) $F_s^+ = F_s^- = 2.37$ pN, (b) $F_s^+ = F_s^- = 0.83$ pN, (c) (d) $F_s^+ = 2.37$ pN, $F_s^- = 0$ pN.